\documentclass[aip,reprint]{revtex4-1}
\usepackage{units}
\usepackage{amsmath}
\usepackage{amssymb}
\usepackage{graphicx}
\usepackage{textcomp}
\usepackage[per-mode=symbol]{siunitx}
\usepackage{natbib}
\usepackage{wasysym}

\usepackage{color}

\begin{document}

\title{Hybrid waveguide-bulk multi-path interferometer with switchable amplitude and phase}

\author{Robert Keil}
\email{robert.keil@uibk.ac.at}
\affiliation{Institut f\"ur Experimentalphysik, Universit\"at Innsbruck, Technikerstra\ss{}e 25, 6020 Innsbruck, Austria}
\author{Thomas Kaufmann}
\altaffiliation{Now at: Swarovski Optik, Absam, Austria} 
\affiliation{Institut f\"ur Experimentalphysik, Universit\"at Innsbruck, Technikerstra\ss{}e 25, 6020 Innsbruck, Austria}
\author{Thomas Kauten}
\affiliation{Institut f\"ur Experimentalphysik, Universit\"at Innsbruck, Technikerstra\ss{}e 25, 6020 Innsbruck, Austria}
\author{Sebastian Gstir}
\affiliation{Institut f\"ur Experimentalphysik, Universit\"at Innsbruck, Technikerstra\ss{}e 25, 6020 Innsbruck, Austria}
\author{Christoph Dittel}
\affiliation{Institut f\"ur Experimentalphysik, Universit\"at Innsbruck, Technikerstra\ss{}e 25, 6020 Innsbruck, Austria}
\author{Ren\'{e} Heilmann}
\affiliation{Institute of Applied Physics, Abbe Center of Photonics, Friedrich-Schiller-Universit\"at Jena, Max-Wien-Platz 1, 07743 Jena, Germany}
\author{Alexander Szameit}
\affiliation{Institute of Applied Physics, Abbe Center of Photonics, Friedrich-Schiller-Universit\"at Jena, Max-Wien-Platz 1, 07743 Jena, Germany}
\author{Gregor Weihs}
\affiliation{Institut f\"ur Experimentalphysik, Universit\"at Innsbruck, Technikerstra\ss{}e 25, 6020 Innsbruck, Austria}

\date{\today}

\begin{abstract}
We design and realise a hybrid interferometer consisting of three paths based on integrated as well as on bulk optical components. This hybrid construction offers a good compromise between stability and footprint on one side and means of intervention on the other. As experimentally verified by the absence of higher-order interferences, amplitude and phase can be manipulated in all paths independently. In conjunction with single photons, the setup can, therefore, be applied for fundamental investigations on quantum mechanics.
\end{abstract}

\maketitle

Interferometers comprised of multiple modes are known to enable a higher precision of phase estimation than in the paradigmatic two-mode scenario \cite{Zernike:PrecisionMeasurementMultiSlit,DAriano:PhasePrecisionMultiPath}. This benefit is often made use of in matter wave interferometry \cite{Weitz:FiveBeamAtomInterferometer,Petrovic:MultiPathAtomChip}. 
In quantum optics, several proposed schemes promise enhancements of non-classical visibilities \cite{Greenberger:TwovsThreeParticleInterference}, improved phase resolution \cite{Spagnolo:MultiPathQuantumInterferometry,Humphreys:MultiPhaseEstimation,Ciampini:QuantumEnhancedMultiparameterEstimation} or better resilience to photon loss \cite{Cooper:MultiPathLossResilience}. In order to exhaust their full potential for precise measurements as well as for high fidelity of state manipulation, a good stability of the interferometer is paramount. Bulk-optical interferometers and solutions based on separate fibers \cite{Weihs:MultiPathInterferometerFiber} are, therefore, at a natural disadvantage compared to waveguide interferometers \cite{Chaboyer:TunableThreePath}. Moreover, such integrated solutions do not suffer from a reduction in interference contrast due to limited mode alignment at the recombining beam splitter. For many applications, however, it is also desirable to be able to switch and manipulate the interferometer. For example, multi-path interferometers with switchable transmission amplitudes are a promising platform for investigations on the foundations of quantum mechanics \cite{Peres:QuaternionQM1979,sorkin1994,sinha10,sollner12,kauten15}, whereas tunable phases can be used to tailor the output state of the photons \cite{Chaboyer:TunableThreePath}. 
The two requirements of stability and capability for manipulation are often in conflict with each other. For example, amplitude switching can be realised in purely integrated settings via nested phase-tunable interferometers at the expense of requiring a substantial amount of independent phase shifters for multi-path interferometers. This is feasible, albeit challenging, in planar silica-on-silicon circuits \cite{Metcalf:MultiPhotonInterferenceMultiport,Carolan:UniversalLinearOptics}, but well beyond the state of the art in three-dimensional fused silica waveguides \cite{Chaboyer:TunableThreePath,Vergyris:OnChipHeraldedPairs}.

A possible trade-off between these goals, in principle applicable to any platform, is offered by hybrid solutions, in which parts of the interferometer are integrated, but the manipulation is performed in free space. One approach is the external modulation of the excitation profile sent into a multi-core fiber, effectively amounting to external, tunable beam splitters \cite{Lee:MCFSLM}. Here, we present an alternative concept: A hybrid three-path Michelson interferometer, which features a fully integrated waveguide beam splitter and, thereby, intrinsically perfect mode overlap after recombination. Therefore, it overcomes limitations on the interference contrast encountered in bulk systems\cite{sollner12}. Amplitude switching is realised by external micro-mirrors mounted on translation stages, such that the reflection coefficients in all interferometer arms can be modulated independently. The phase is controlled by fine tuning of the mirror position. We characterise this interferometer by measuring the contrast of two-path interference and by testing for undesired
systematic correlations between the transmission amplitudes of the individual mirror settings of all three paths. The latter test is performed by measuring so-called higher-order interferences \cite{sinha10}, which are expected to vanish in the regime of classical electrodynamics \cite{kastner16}. No higher-order interferences are detected, suggesting that the interferometer is free of significant systematic correlations.

\begin{figure}
 \centering
 \includegraphics[width=\columnwidth]{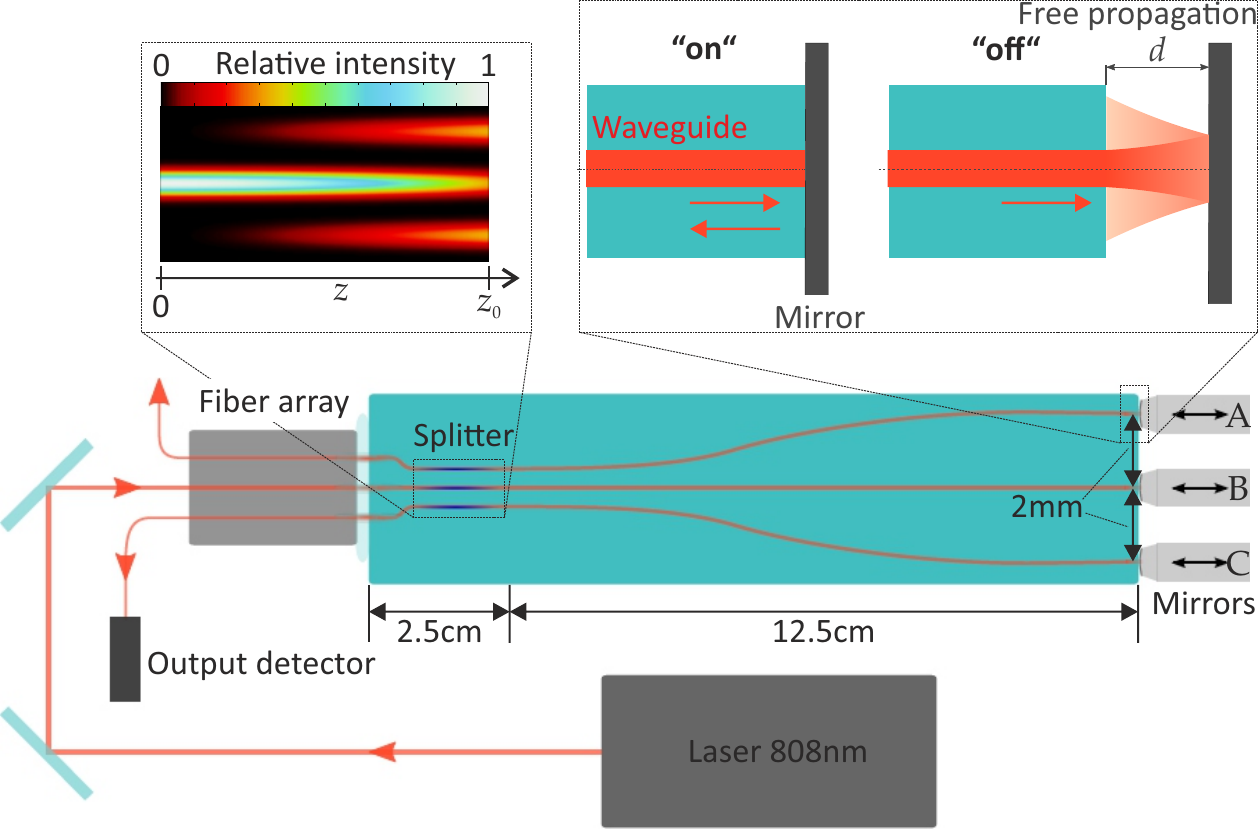}
 \caption{\label{fig:setup} Hybrid three-path interferometer. Coherent laser light is coupled into a waveguide splitter, distributing the light evenly via evanescent coupling (left inset). Mirrors at the opposite end of the chip reflect the light back to the splitter, forming a Michelson-interferometer. Longitudinal displacement of the mirrors is used to enable or suppress reflection (right inset). The waveguides were fabricated in a single inscription run within the splitter (drawn in blue), which facilitates evanescent coupling, whereas multiple inscription runs were used in the curved parts (red), a fabrication regime which reduces bending losses.}
\end{figure}

A schematic drawing of the interferometer can be seen in Fig. \ref{fig:setup}.  The integrated part of the setup consists of three waveguides, individually accessible at the input face via a polarisation maintaining fiber array with \SI{127}{\micro\meter} pitch. The input light ($\lambda=\SI{808}{\nano\meter}$) is linearly polarised in the chip plane and power stabilised to a relative rms-noise of $0.35\%$ over the entire duration of the experiment. The three-path beamsplitter is a linear array of the three guides with a design length $z_0=\arctan\left(\sqrt{2}\right)/\left(\sqrt{2}C\right)$ for the nearest-neighbour evanescent coupling rate $C$. This configuration leads to an equal splitting of light power when the central site is excited \cite{Graefe:2DSimulatorBiphotonWalk,Perez-Leija2013} (see inset). The waveguides were inscribed $\SI{200}{\micro\meter}$ below the surface in fused silica glass (Corning 7980) by femtosecond laser pulses \cite{Meany:LaserWrittenQuantumCircuitsReview} (wavelength $\SI{515}{\nano\meter}$, repetition rate $\SI{200}{\kilo\hertz}$, pulse duration $\SI{270}{\femto\second}$, inscription velocity $\SI{4.2}{\milli\meter/\second}$). Here, the effective coupling parameter is $C=\SI{0.56}{\centi\meter^{-1}}$, associated with a distance of $\SI{26}{\micro\meter}$ between the waveguides. One of the outer waveguides acts as output port for the reflected signals, which are registered by a biased $\mathrm{Si}$ detector. Note that the splitter is not balanced for light arriving in one of its outer ports. Therefore, the signals reflected in the three paths will contribute to the output with different magnitudes. In order to enable interfacing to macroscopic mirrors in the free-space part of the setup the distance between the waveguides is adiabatically increased to \SI{2}{\milli\meter} towards the opposite end of the chip. Low bending losses are ensured by increasing the refractive index contrast of the waveguides via fourfold inscription of their track in all curved segments. This leads to a tighter confinement of the modes, thereby, also suppressing undesired coupling before and after the splitter \cite{Heilmann:AdiabaticWriting}. The effective propagation loss in all waveguides, averaged over the whole length of the chip, was measured to be $\SI{0.15\pm 0.05}{\decibel/\centi\meter}$.

\begin{figure}
 \centering
 \includegraphics[width=\columnwidth]{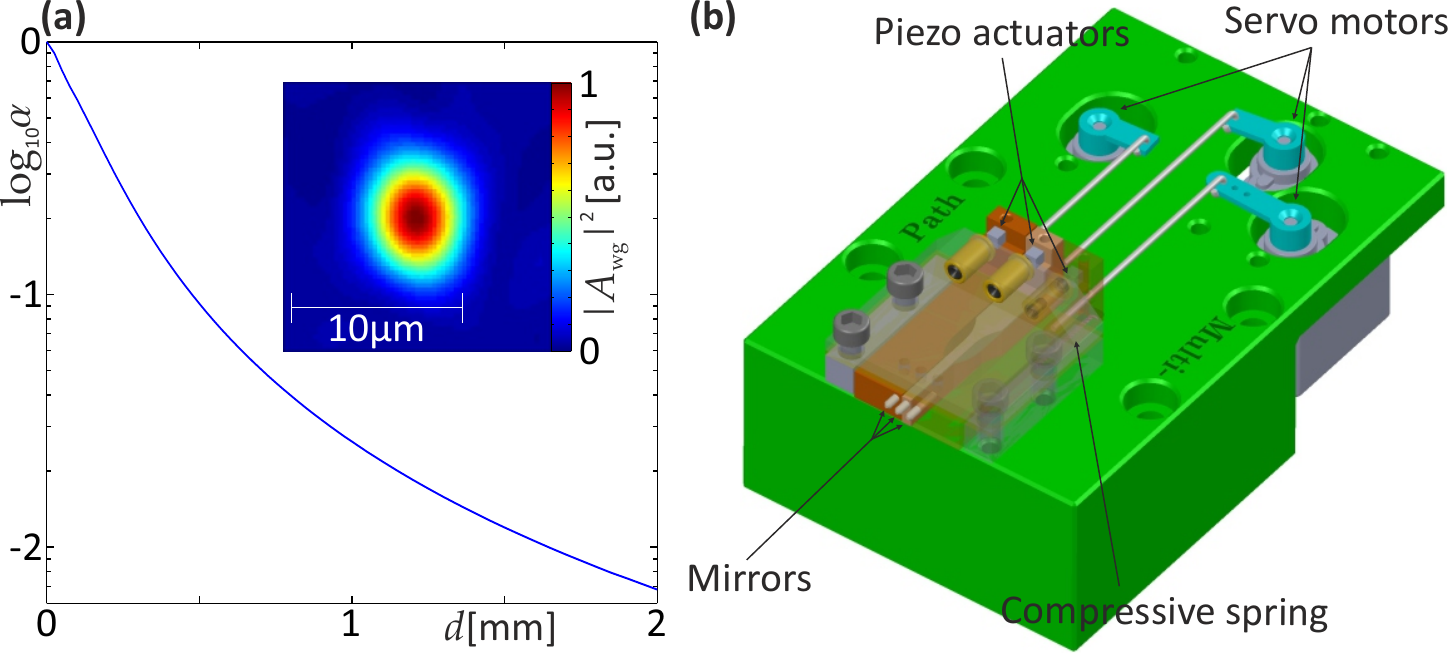}
 \caption{\label{fig:mirror} Mirror setup. \textbf{(a)} Numerical simulation of the back-coupling efficiency $\alpha$ as defined in Eq.~(\ref{overlap}) in dependence of the chip-mirror distance $d$. The inset shows the measured mode intensity of the waveguide at $\lambda=\SI{808}{\nano\meter}$. \textbf{(b)} Illustration of the mirror positioning setup. Servo motors pull the mirrors away from the chip and compress springs, which push the mirrors back into the proximity position ("on"-state) when released. Piezo actuators can be used for fine adjustment.}
\end{figure}

The mirrors of this three-path Michelson interferometer are realised externally. Ceramic fiber ferrules with a flat surface, \SI{0.6}{\milli\meter} in diameter at their tip were dielectrically coated with layers of $\mathrm{TiO}_2$ and $\mathrm{SiO}_2$, reaching a reflectivity of $99.9\%$ at $\lambda$. The operating principle of the switchable mirrors is illustrated in the right inset of Fig.~\ref{fig:setup}: In the ``on''-state the ferrule is placed very close and parallel to the chip end facet, such that practically no diffraction occurs and the light can be efficiently coupled back into the waveguide. In the ``off''-state, however, the distance $d$ is increased, such that diffraction in the gap prevents an efficient recoupling. To prevent Fresnel reflections at the chip surface, which would lead to additional interference, the gap between chip and mirror is filled with refractive index matching gel ($n=1.45$). The recoupling efficiency is then solely determined by the overlap between the waveguide mode field $A_{\mathrm{wg}}$ and the light field after free propagation over the distance $2d$ through the gel medium, $A_{2d}$ (see, e.g., Ref.~\onlinecite{Hunsperger:IntegratedOpticsBook}): 
\begin{equation}
\alpha=\frac{\left|\iint A_{\mathrm{wg}}(x,y)A_{2d}^{\ast}(x,y)\mathrm{d}x\mathrm{d}y\right|^2}{\iint \left|A_{\mathrm{wg}}(x,y)\right|^2\mathrm{d}x\mathrm{d}y\iint \left|A_{2d}(x,y)\right|^2\mathrm{d}x\mathrm{d}y},
\label{overlap}
\end{equation} 
with transverse coordinates $x$ and $y$. 
Fig.~\ref{fig:mirror}\textbf{(a)} shows a numerical calculation of this overlap for a diffraction of the measured waveguide mode field within the paraxial approximation (which is well met in this case). In the following experiments we have $d\approx \SI{2}{\milli\meter}$, which implies that the reflected signal will be reduced by about two orders of magnitude in the ``off''-state. Practically, even larger switching contrasts can be expected, as the two surfaces will not be perfectly parallel, causing a lateral offset of the reflected beam increasing with $d$. The mechanical mounting of the mirrors is illustrated in Fig.~\ref{fig:mirror}\textbf{(b)}, with electromechanical components for coarse positioning and piezoelectric actuators for fine adjustment. All mirrors can be moved independently from one another.

To stabilise the interferometer against acoustic distortions, the waveguide chip is embedded in an aluminium housing. For passive and active thermal stabilisation, the whole setup including the mirror stage and the input/output fiber interface is shielded with polystyrene sheets and the temperature is regulated by a PI controller connected to heating mats, reducing long-term temperature fluctuations to about $\SI{5}{\milli\kelvin}$. In the experiment, we run cycles with all eight combinations of mirror settings, measuring photovoltages at the detector: We denote $P_0$ as the voltage obtained for all mirrors in the ``off''-state, $P_{\mathrm{A}}$ for only mirror A being in the ``on''-state, $P_{\mathrm{AB}}$ for mirrors A and B in the ``on''-state, but C in the ``off''-state, etc. The order of these settings is permuted randomly from cycle to cycle to eliminate influences of drift processes on the relative distribution of the voltages. For all settings, the detector averages over \SI{30}{\second} of signal and an average idle time of \SI{3}{\minute} was introduced after each switching to account for mechanical inertia in the setup, likely arising from the mirror mounts.

\begin{figure}
 \centering
 \includegraphics[width=\columnwidth]{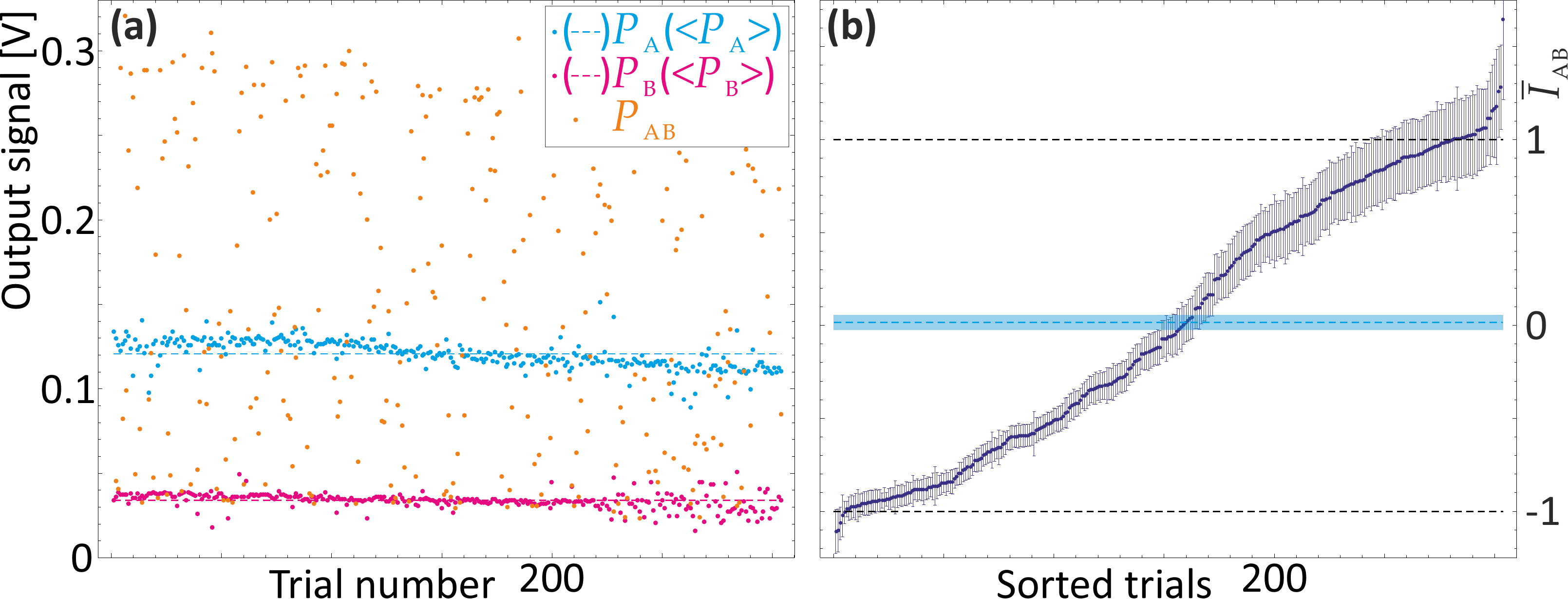}
 \caption{\label{fig:ABTerm}Two-path interference. \textbf{(a)} Output power series for 304 cycles of mirror settings. Shown are the signals for settings with only mirror A reflecting (cyan), only mirror B (magenta) and mirrors A and B simultaneously (orange). The horizontal lines indicate the averages of the two single-path settings. \textbf{(b)} Normalised interference term $\bar{I}_{\mathrm{AB}}$, sorted in ascending order and covering the whole range between fully constructive (line at $\bar{I}_{\mathrm{AB}}=1$) and fully destructive interference ($\bar{I}_{\mathrm{AB}}=-1$). The error bars indicate one standard deviation assuming independent Gaussian fluctuations of the underlying single-path rates with a magnitude obtained from \textbf{(a)}. The blue horizontal line indicates the mean of the distribution and the shaded bar its uncertainty.}
\end{figure}

In order to characterise the coherence of the interferometer, one can exploit the fact that repositioning the mirrors without piezo control is sufficiently coarse to randomise the phase. Therefore, a repeated switching of the mirror settings allows to explore the entire phase space of the interferometer. In particular, the two-path interference should cover the whole range between fully constructive and fully destructive interference, i.e., for paths A and B $\left(\sqrt{P_{\mathrm{A}}}-\sqrt{P_{\mathrm{B}}}\right)^2\leq P_{\mathrm{AB}}\leq\left(\sqrt{P_{\mathrm{A}}}+\sqrt{P_{\mathrm{B}}}\right)^2$. Reductions in this interference contrast can then be attributed to a deteriorated coherence. In total, $322$ measurement cycles were performed within 164 hours. The single-path data resulting from terms $P_{\mathrm{A}},P_{\mathrm{B}},P_{\mathrm{C}}$ was filtered for outliers (as in the absence of interference one can confidently assign values strongly deviating from the mean to mechanical switching failure, which occur with $\apprle 2\%$ probability in this setup) via Grubb's method\cite{grubbs69} (significance level $95\%$), leaving a data set of $304$ complete trials for further analysis. Fig.~\ref{fig:ABTerm}\textbf{(a)} shows this data for the interference between paths A and B. The two single path-terms (cyan and magenta) decrease only slowly during the duration of the experiment (relative drift $\approx\SI{-0.07}{\%/\hour}$), indicating the good long-term stability of the setup. The two-path term $P_{\mathrm{AB}}$ is distributed over a wide interval of values (orange points). From the difference between two-path and single-path measurements one can reconstruct the interference term 
\begin{equation}
I_{\mathrm{AB}}\equiv P_{\mathrm{AB}}-P_{\mathrm{A}}-P_{\mathrm{B}}
\label{twopathinterference}
\end{equation}
for all cycles. Its normalised version $\bar{I}_{\mathrm{AB}}\equiv I_{\mathrm{AB}}/\left(2\sqrt{P_{\mathrm{A}}P_{\mathrm{B}}}\right)$ (the cosine of the phase difference between the two paths) is plotted in Fig.~\ref{fig:ABTerm}\textbf{(b)}, demonstrating that both extremes of fully constructive and fully destructive interference are reached. For a fully coherent process and a uniform distribution of phases one would expect $\left\langle \bar{I}_{\mathrm{AB}}\right\rangle=0$ and $\sigma_{\bar{I}_{\mathrm{AB}}}=1/\sqrt{2}$ for the mean and standard deviation of the interference terms. Here, one finds from the data $\left\langle \bar{I}_{\mathrm{AB}}\right\rangle\approx 0.02\pm 0.04$ and $\sigma_{\bar{I}_{\mathrm{AB}}}\approx \left(1.01\pm 0.04\right)/\sqrt{2}$, with the error intervals corresponding to one standard error of mean and standard deviation, respectively. This is in full accordance with the aforementioned expectations and one can, therefore, conclude that a degree of coherence near unity is reached between these paths.

\begin{figure}
 \centering
 \includegraphics[width=\columnwidth]{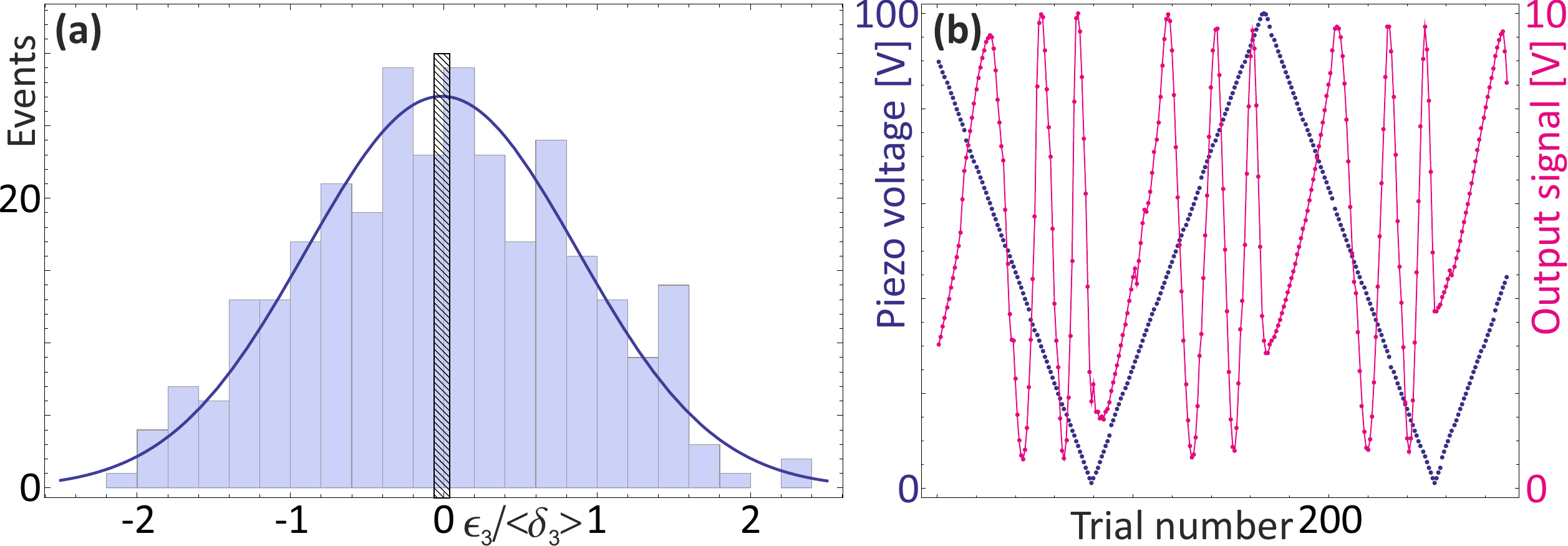}
 \caption{\label{fig:Histogram} \textbf{(a)} Three-path interference. Distribution of the $304$ measured second-order interference terms $\epsilon_3$, normalised to the total two-path interference $\delta_3$. The curve shows a normal distribution with mean and variance according to the measured data. The shaded bar indicates the mean of the distribution and its uncertainty. \textbf{(b)} Phase scan. A periodic voltage ramp (blue) sweeps the phase of the two-path interference between channels A and B, leading to fringes in the output signal (magenta).}
\end{figure}

In a next step, we let all three paths contribute to the interference. One can show straightforwardly from Poynting's theorem relating power and field amplitude of transverse waves $P\propto\left|E\right|^2$ and the superposition principle that interference in such a multi-path interferometer must occur in pairs of paths, that is:
\begin{equation}
\begin{aligned}
\left|E_\mathrm{A}+E_\mathrm{B}+E_\mathrm{C}\right|^2&\propto P_{\mathrm{ABC}}\nonumber\\
&=P_{\mathrm{A}}+P_{\mathrm{B}}+P_{\mathrm{C}}+I_{\mathrm{AB}}+I_{\mathrm{AC}}+I_{\mathrm{BC}},\nonumber
\end{aligned}
\end{equation}
with two-path interference terms, as defined in~(\ref{twopathinterference}). Taking into account the background voltage $P_0$, the following quantity must vanish:
\begin{equation}
\epsilon_3\equiv P_{\mathrm{ABC}}-P_{\mathrm{AB}}-P_{\mathrm{AC}}-P_{\mathrm{BC}}+P_{\mathrm{A}}+P_{\mathrm{B}}+P_{\mathrm{C}}-P_{0}=0.
\nonumber
\end{equation}
In the context of quantum mechanics and its possible extensions, $\epsilon_3$ is referred to as second-order interference and can be used as a probe to test generalised interference theories \cite{sorkin1994,sinha10,park12,sollner12,kauten15}. Here, the absence of $\epsilon_3$ is implied by electrodynamics\cite{kastner16} and will be used for a consistency check of the interferometer. Note that $\epsilon_3=0$ must hold also in the presence of non-vanishing reflected signals in the ``off''-state of a mirror. We calculate $\epsilon_3$ for all measurement cycles and normalise it by the sum of all regular two-path interference terms $\delta_3=\left|I_{\mathrm{AB}}\right|+\left|I_{\mathrm{AC}}\right|+\left|I_{\mathrm{BC}}\right|$ to produce a quantity which is independent of input power and coherence. Note that individual measurement results can produce large deviations from zero, due to the random phases of the interferometer in the individual mirror settings. One can show from theoretical considerations, however, that the expectation value of $\epsilon_3$ stays zero, regardless of the magnitude of such phase fluctuations. The only notable exception to this rule would be produced by correlations between the reflectivity and phase of a mirror with the settings of the other mirrors, as such correlations would lead to cross-talk between the paths of the interferometer and bias the second-order interference term \cite{kauten15}. The measured distribution of $\epsilon_3$ is plotted in Fig.~\ref{fig:Histogram}\textbf{(a)}. Indeed, one obtains a normalised mean value of $\left\langle\epsilon_3\right\rangle/\left\langle\delta_3\right\rangle=-0.01\pm 0.05$ (the uncertainty being one standard error of mean) from the experimental data. This strongly suggests that the reflected signals from all paths interfere in a consistent manner and no significant cross-talk exists. 

Finally, we demonstrate the phase-tunability in the hybrid interferometer. To this end, mirrors A and B are brought into the ``on"-state and the piezo crystal on mirror A is subjected to a linear and periodic voltage ramp between 0 and \SI{100}{\volt} (blue curve in Fig.~\ref{fig:Histogram}\textbf{(b)}), sweeping the relative phase between interferometer arms A and B. The measured output signal $P_{\mathrm{AB}}$ (magenta) follows this drive and oscillates between constructive and destructive interference. The whole voltage range corresponds to about 2.5 periods of oscillation, equivalent to $1.25 \lambda/n\approx\SI{700}{\nano\meter}$ mirror displacement. This shows that any point in phase space can be reached by the piezo stacks in a repeatable way. The slower response immediately after the extrema of the voltage curve, where the piezo crystal expansion switches direction, arises from hysteresis of the piezo and is amplified by the aforementioned mechanical moments of inertia. In experiments requiring stabilisation around a single phase one would avoid this issue by picking a mid-range voltage and, if necessary, using closed loop operation.

In this work we presented a hybrid freespace-integrated multi-path interferometer. The beam splitter of this Michelson interferometer and practically the entire optical path length are embedded in intrinsically stable and alignment-free optical waveguides, whereas only the mirrors are implemented with bulk optics. The setup provides full control over path transmissions and phases with good long-term stability and a high degree of coherence. The absence of significant second-order interference is strong evidence for the mutual independence of the path settings. Vice versa, when operated with heralded single photons, the setup can be applied towards fundamental tests of quantum mechanics in a much more compact setting than bulk interferometers. While being free of systematic errors on the current level of precision, that precision still falls short of the state of the art \cite{kauten15}. 
A large part of this imprecision arises from the fact that each amplitude switching by mirror translation also affects the relative phase between the paths. An improved version of the hybrid interferometer would, therefore, aim at decoupling transmission and phase switching from each other. One possibility would be a Mach-Zehnder configuration with cavities, e.g. etched into the glass after laser structuring \cite{Marcinkevicius:LaserChannelEtching,Crespi:MeasuringProteinConc_EntPhot}, into which shutters and phase shifters can be inserted separately. This would have the additional benefit of smaller moving masses and, therefore, higher feasible switching rates. As an upgrade to the present system, one could further use micro-scanning mirrors for faster operation. Due to shot noise in the single photon regime, the precision will also depend on the signal throughput. This can be improved by operating in the telecom band with superconducting detectors \cite{Pernice:SCNanowiresOnChip} or utilizing materials with higher integration density and, therefore, shorter propagation lengths \cite{Metcalf:MultiPhotonInterferenceMultiport,Carolan:UniversalLinearOptics}. Note that detector nonlinearities as studied in Ref.~\onlinecite{kauten14} bias the result in the order $10^{-6}$, hence playing no role at the present level of precision, will eventually have to be taken into account in future experiments.  Finally, owing to their good coherence properties, hybrid multi-path interferometers with improved precision may also be used to test for hypercomplex generalisations of quantum theory \cite{Peres:QuaternionQM1979,Procopio:SagnacHypercomplexTest}.

The authors acknowledge support by the Austrian Science Fund (projects I 2562-N27 and M 1849), the Canadian Institute for Advanced Research (Quantum Information Science Program), the European Research Council (project 257531 - EnSeNa) and the German Ministry of Education and Research (Center for Innovation Competence program, grant 03Z1HN31).


\begin{thebibliography}{33}%
\makeatletter
\providecommand \@ifxundefined [1]{%
 \@ifx{#1\undefined}
}%
\providecommand \@ifnum [1]{%
 \ifnum #1\expandafter \@firstoftwo
 \else \expandafter \@secondoftwo
 \fi
}%
\providecommand \@ifx [1]{%
 \ifx #1\expandafter \@firstoftwo
 \else \expandafter \@secondoftwo
 \fi
}%
\providecommand \natexlab [1]{#1}%
\providecommand \enquote  [1]{``#1''}%
\providecommand \bibnamefont  [1]{#1}%
\providecommand \bibfnamefont [1]{#1}%
\providecommand \citenamefont [1]{#1}%
\providecommand \href@noop [0]{\@secondoftwo}%
\providecommand \href [0]{\begingroup \@sanitize@url \@href}%
\providecommand \@href[1]{\@@startlink{#1}\@@href}%
\providecommand \@@href[1]{\endgroup#1\@@endlink}%
\providecommand \@sanitize@url [0]{\catcode `\\12\catcode `\$12\catcode
  `\&12\catcode `\#12\catcode `\^12\catcode `\_12\catcode `\%12\relax}%
\providecommand \@@startlink[1]{}%
\providecommand \@@endlink[0]{}%
\providecommand \url  [0]{\begingroup\@sanitize@url \@url }%
\providecommand \@url [1]{\endgroup\@href {#1}{\urlprefix }}%
\providecommand \urlprefix  [0]{URL }%
\providecommand \Eprint [0]{\href }%
\providecommand \doibase [0]{http://dx.doi.org/}%
\providecommand \selectlanguage [0]{\@gobble}%
\providecommand \bibinfo  [0]{\@secondoftwo}%
\providecommand \bibfield  [0]{\@secondoftwo}%
\providecommand \translation [1]{[#1]}%
\providecommand \BibitemOpen [0]{}%
\providecommand \bibitemStop [0]{}%
\providecommand \bibitemNoStop [0]{.\EOS\space}%
\providecommand \EOS [0]{\spacefactor3000\relax}%
\providecommand \BibitemShut  [1]{\csname bibitem#1\endcsname}%
\let\auto@bib@innerbib\@empty
\bibitem [{\citenamefont
  {Zernike}(1950)}]{Zernike:PrecisionMeasurementMultiSlit}%
  \BibitemOpen
  \bibfield  {author} {\bibinfo {author} {\bibfnamefont {F.}~\bibnamefont
  {Zernike}},\ }\href {\doibase 10.1364/JOSA.40.000326} {\bibfield  {journal}
  {\bibinfo  {journal} {J. Opt. Soc. Am.}\ }\textbf {\bibinfo {volume} {40}},\
  \bibinfo {pages} {326} (\bibinfo {year} {1950})}\BibitemShut {NoStop}%
\bibitem [{\citenamefont {D'Ariano}\ and\ \citenamefont
  {Paris}(1997)}]{DAriano:PhasePrecisionMultiPath}%
  \BibitemOpen
  \bibfield  {author} {\bibinfo {author} {\bibfnamefont {G.~M.}\ \bibnamefont
  {D'Ariano}}\ and\ \bibinfo {author} {\bibfnamefont {M.~G.~A.}\ \bibnamefont
  {Paris}},\ }\href {\doibase 10.1103/PhysRevA.55.2267} {\bibfield  {journal}
  {\bibinfo  {journal} {Phys. Rev. A}\ }\textbf {\bibinfo {volume} {55}},\
  \bibinfo {pages} {2267} (\bibinfo {year} {1997})}\BibitemShut {NoStop}%
\bibitem [{\citenamefont {Weitz}, \citenamefont {Heupel},\ and\ \citenamefont
  {H\"ansch}(1996)}]{Weitz:FiveBeamAtomInterferometer}%
  \BibitemOpen
  \bibfield  {author} {\bibinfo {author} {\bibfnamefont {M.}~\bibnamefont
  {Weitz}}, \bibinfo {author} {\bibfnamefont {T.}~\bibnamefont {Heupel}}, \
  and\ \bibinfo {author} {\bibfnamefont {T.~W.}\ \bibnamefont {H\"ansch}},\
  }\href {\doibase 10.1103/PhysRevLett.77.2356} {\bibfield  {journal} {\bibinfo
   {journal} {Phys. Rev. Lett.}\ }\textbf {\bibinfo {volume} {77}},\ \bibinfo
  {pages} {2356} (\bibinfo {year} {1996})}\BibitemShut {NoStop}%
\bibitem [{\citenamefont {Petrovic}\ \emph {et~al.}(2013)\citenamefont
  {Petrovic}, \citenamefont {Herrera}, \citenamefont {Lombardi}, \citenamefont
  {Sch\"{a}fer},\ and\ \citenamefont
  {Cataliotti}}]{Petrovic:MultiPathAtomChip}%
  \BibitemOpen
  \bibfield  {author} {\bibinfo {author} {\bibfnamefont {J.}~\bibnamefont
  {Petrovic}}, \bibinfo {author} {\bibfnamefont {I.}~\bibnamefont {Herrera}},
  \bibinfo {author} {\bibfnamefont {P.}~\bibnamefont {Lombardi}}, \bibinfo
  {author} {\bibfnamefont {F.}~\bibnamefont {Sch\"{a}fer}}, \ and\ \bibinfo
  {author} {\bibfnamefont {F.~S.}\ \bibnamefont {Cataliotti}},\ }\href
  {http://stacks.iop.org/1367-2630/15/i=4/a=043002} {\bibfield  {journal}
  {\bibinfo  {journal} {New J. Phys.}\ }\textbf {\bibinfo {volume} {15}},\
  \bibinfo {pages} {043002} (\bibinfo {year} {2013})}\BibitemShut {NoStop}%
\bibitem [{\citenamefont {Greenberger}, \citenamefont {Horne},\ and\
  \citenamefont
  {Zeilinger}(2000)}]{Greenberger:TwovsThreeParticleInterference}%
  \BibitemOpen
  \bibfield  {author} {\bibinfo {author} {\bibfnamefont {D.~M.}\ \bibnamefont
  {Greenberger}}, \bibinfo {author} {\bibfnamefont {M.}~\bibnamefont {Horne}},
  \ and\ \bibinfo {author} {\bibfnamefont {A.}~\bibnamefont {Zeilinger}},\
  }\href {\doibase
  10.1002/(SICI)1521-3978(200004)48:4<243::AID-PROP243>3.0.CO;2-7} {\bibfield
  {journal} {\bibinfo  {journal} {Fortschr. Phys.}\ }\textbf {\bibinfo {volume}
  {48}},\ \bibinfo {pages} {243} (\bibinfo {year} {2000})}\BibitemShut
  {NoStop}%
\bibitem [{\citenamefont {Spagnolo}\ \emph {et~al.}(2012)\citenamefont
  {Spagnolo}, \citenamefont {Aparo}, \citenamefont {Vitelli}, \citenamefont
  {Crespi}, \citenamefont {Ramponi}, \citenamefont {Osellame}, \citenamefont
  {Mataloni},\ and\ \citenamefont
  {Sciarrino}}]{Spagnolo:MultiPathQuantumInterferometry}%
  \BibitemOpen
  \bibfield  {author} {\bibinfo {author} {\bibfnamefont {N.}~\bibnamefont
  {Spagnolo}}, \bibinfo {author} {\bibfnamefont {L.}~\bibnamefont {Aparo}},
  \bibinfo {author} {\bibfnamefont {C.}~\bibnamefont {Vitelli}}, \bibinfo
  {author} {\bibfnamefont {A.}~\bibnamefont {Crespi}}, \bibinfo {author}
  {\bibfnamefont {R.}~\bibnamefont {Ramponi}}, \bibinfo {author} {\bibfnamefont
  {R.}~\bibnamefont {Osellame}}, \bibinfo {author} {\bibfnamefont
  {P.}~\bibnamefont {Mataloni}}, \ and\ \bibinfo {author} {\bibfnamefont
  {F.}~\bibnamefont {Sciarrino}},\ }\href {http://dx.doi.org/10.1038/srep00862}
  {\bibfield  {journal} {\bibinfo  {journal} {Sci. Rep.}\ }\textbf {\bibinfo
  {volume} {2}},\ \bibinfo {pages} {862} (\bibinfo {year} {2012})}\BibitemShut
  {NoStop}%
\bibitem [{\citenamefont {Humphreys}\ \emph {et~al.}(2013)\citenamefont
  {Humphreys}, \citenamefont {Barbieri}, \citenamefont {Datta},\ and\
  \citenamefont {Walmsley}}]{Humphreys:MultiPhaseEstimation}%
  \BibitemOpen
  \bibfield  {author} {\bibinfo {author} {\bibfnamefont {P.~C.}\ \bibnamefont
  {Humphreys}}, \bibinfo {author} {\bibfnamefont {M.}~\bibnamefont {Barbieri}},
  \bibinfo {author} {\bibfnamefont {A.}~\bibnamefont {Datta}}, \ and\ \bibinfo
  {author} {\bibfnamefont {I.~A.}\ \bibnamefont {Walmsley}},\ }\href {\doibase
  10.1103/PhysRevLett.111.070403} {\bibfield  {journal} {\bibinfo  {journal}
  {Phys. Rev. Lett.}\ }\textbf {\bibinfo {volume} {111}},\ \bibinfo {pages}
  {070403} (\bibinfo {year} {2013})}\BibitemShut {NoStop}%
\bibitem [{\citenamefont {Ciampini}\ \emph {et~al.}(2016)\citenamefont
  {Ciampini}, \citenamefont {Spagnolo}, \citenamefont {Vitelli}, \citenamefont
  {Pezz\'{e}}, \citenamefont {Smerzi},\ and\ \citenamefont
  {Sciarrino}}]{Ciampini:QuantumEnhancedMultiparameterEstimation}%
  \BibitemOpen
  \bibfield  {author} {\bibinfo {author} {\bibfnamefont {M.~A.}\ \bibnamefont
  {Ciampini}}, \bibinfo {author} {\bibfnamefont {N.}~\bibnamefont {Spagnolo}},
  \bibinfo {author} {\bibfnamefont {C.}~\bibnamefont {Vitelli}}, \bibinfo
  {author} {\bibfnamefont {L.}~\bibnamefont {Pezz\'{e}}}, \bibinfo {author}
  {\bibfnamefont {A.}~\bibnamefont {Smerzi}}, \ and\ \bibinfo {author}
  {\bibfnamefont {F.}~\bibnamefont {Sciarrino}},\ }\href
  {http://dx.doi.org/10.1038/srep28881} {\bibfield  {journal} {\bibinfo
  {journal} {Sci. Rep.}\ }\textbf {\bibinfo {volume} {6}},\ \bibinfo {pages}
  {28881} (\bibinfo {year} {2016})}\BibitemShut {NoStop}%
\bibitem [{\citenamefont {Cooper}\ and\ \citenamefont
  {Dunningham}(2011)}]{Cooper:MultiPathLossResilience}%
  \BibitemOpen
  \bibfield  {author} {\bibinfo {author} {\bibfnamefont {J.~J.}\ \bibnamefont
  {Cooper}}\ and\ \bibinfo {author} {\bibfnamefont {J.~A.}\ \bibnamefont
  {Dunningham}},\ }\href {http://stacks.iop.org/1367-2630/13/i=11/a=115003}
  {\bibfield  {journal} {\bibinfo  {journal} {New J. Phys.}\ }\textbf {\bibinfo
  {volume} {13}},\ \bibinfo {pages} {115003} (\bibinfo {year}
  {2011})}\BibitemShut {NoStop}%
\bibitem [{\citenamefont {Weihs}\ \emph {et~al.}(1996)\citenamefont {Weihs},
  \citenamefont {Reck}, \citenamefont {Weinfurter},\ and\ \citenamefont
  {Zeilinger}}]{Weihs:MultiPathInterferometerFiber}%
  \BibitemOpen
  \bibfield  {author} {\bibinfo {author} {\bibfnamefont {G.}~\bibnamefont
  {Weihs}}, \bibinfo {author} {\bibfnamefont {M.}~\bibnamefont {Reck}},
  \bibinfo {author} {\bibfnamefont {H.}~\bibnamefont {Weinfurter}}, \ and\
  \bibinfo {author} {\bibfnamefont {A.}~\bibnamefont {Zeilinger}},\ }\href
  {\doibase 10.1364/OL.21.000302} {\bibfield  {journal} {\bibinfo  {journal}
  {Opt. Lett.}\ }\textbf {\bibinfo {volume} {21}},\ \bibinfo {pages} {302}
  (\bibinfo {year} {1996})}\BibitemShut {NoStop}%
\bibitem [{\citenamefont {Chaboyer}\ \emph {et~al.}(2015)\citenamefont
  {Chaboyer}, \citenamefont {Meany}, \citenamefont {Helt}, \citenamefont
  {Withford},\ and\ \citenamefont {Steel}}]{Chaboyer:TunableThreePath}%
  \BibitemOpen
  \bibfield  {author} {\bibinfo {author} {\bibfnamefont {Z.}~\bibnamefont
  {Chaboyer}}, \bibinfo {author} {\bibfnamefont {T.}~\bibnamefont {Meany}},
  \bibinfo {author} {\bibfnamefont {L.~G.}\ \bibnamefont {Helt}}, \bibinfo
  {author} {\bibfnamefont {M.~J.}\ \bibnamefont {Withford}}, \ and\ \bibinfo
  {author} {\bibfnamefont {M.~J.}\ \bibnamefont {Steel}},\ }\href
  {http://dx.doi.org/10.1038/srep09601} {\bibfield  {journal} {\bibinfo
  {journal} {Sci. Rep.}\ }\textbf {\bibinfo {volume} {5}},\ \bibinfo {pages}
  {9601} (\bibinfo {year} {2015})}\BibitemShut {NoStop}%
\bibitem [{\citenamefont {Peres}(1979)}]{Peres:QuaternionQM1979}%
  \BibitemOpen
  \bibfield  {author} {\bibinfo {author} {\bibfnamefont {A.}~\bibnamefont
  {Peres}},\ }\href {\doibase 10.1103/PhysRevLett.42.683} {\bibfield  {journal}
  {\bibinfo  {journal} {Phys. Rev. Lett.}\ }\textbf {\bibinfo {volume} {42}},\
  \bibinfo {pages} {683} (\bibinfo {year} {1979})}\BibitemShut {NoStop}%
\bibitem [{\citenamefont {Sorkin}(1994)}]{sorkin1994}%
  \BibitemOpen
  \bibfield  {author} {\bibinfo {author} {\bibfnamefont {R.~D.}\ \bibnamefont
  {Sorkin}},\ }\href {\doibase 10.1142/S021773239400294X} {\bibfield  {journal}
  {\bibinfo  {journal} {Mod. Phys. Lett. A}\ }\textbf {\bibinfo {volume}
  {09}},\ \bibinfo {pages} {3119} (\bibinfo {year} {1994})}\BibitemShut
  {NoStop}%
\bibitem [{\citenamefont {Sinha}\ \emph {et~al.}(2010)\citenamefont {Sinha},
  \citenamefont {Couteau}, \citenamefont {Jennewein}, \citenamefont
  {Laflamme},\ and\ \citenamefont {Weihs}}]{sinha10}%
  \BibitemOpen
  \bibfield  {author} {\bibinfo {author} {\bibfnamefont {U.}~\bibnamefont
  {Sinha}}, \bibinfo {author} {\bibfnamefont {C.}~\bibnamefont {Couteau}},
  \bibinfo {author} {\bibfnamefont {T.}~\bibnamefont {Jennewein}}, \bibinfo
  {author} {\bibfnamefont {R.}~\bibnamefont {Laflamme}}, \ and\ \bibinfo
  {author} {\bibfnamefont {G.}~\bibnamefont {Weihs}},\ }\href {\doibase
  10.1126/science.1190545} {\bibfield  {journal} {\bibinfo  {journal}
  {Science}\ }\textbf {\bibinfo {volume} {329}},\ \bibinfo {pages} {418}
  (\bibinfo {year} {2010})}\BibitemShut {NoStop}%
\bibitem [{\citenamefont {S\"ollner}\ \emph {et~al.}(2012)\citenamefont
  {S\"ollner}, \citenamefont {Gsch\"osser}, \citenamefont {Mai}, \citenamefont
  {Pressl}, \citenamefont {V\"or\"os},\ and\ \citenamefont
  {Weihs}}]{sollner12}%
  \BibitemOpen
  \bibfield  {author} {\bibinfo {author} {\bibfnamefont {I.}~\bibnamefont
  {S\"ollner}}, \bibinfo {author} {\bibfnamefont {B.}~\bibnamefont
  {Gsch\"osser}}, \bibinfo {author} {\bibfnamefont {P.}~\bibnamefont {Mai}},
  \bibinfo {author} {\bibfnamefont {B.}~\bibnamefont {Pressl}}, \bibinfo
  {author} {\bibfnamefont {Z.}~\bibnamefont {V\"or\"os}}, \ and\ \bibinfo
  {author} {\bibfnamefont {G.}~\bibnamefont {Weihs}},\ }\href {\doibase
  10.1007/s10701-011-9597-5} {\bibfield  {journal} {\bibinfo  {journal} {Found.
  Phys.}\ }\textbf {\bibinfo {volume} {42}},\ \bibinfo {pages} {742} (\bibinfo
  {year} {2012})}\BibitemShut {NoStop}%
\bibitem [{\citenamefont {Kauten}\ \emph {et~al.}(2015)\citenamefont {Kauten},
  \citenamefont {Keil}, \citenamefont {Kaufmann}, \citenamefont {Pressl},\ and\
  \citenamefont {Weihs}}]{kauten15}%
  \BibitemOpen
  \bibfield  {author} {\bibinfo {author} {\bibfnamefont {T.}~\bibnamefont
  {Kauten}}, \bibinfo {author} {\bibfnamefont {R.}~\bibnamefont {Keil}},
  \bibinfo {author} {\bibfnamefont {T.}~\bibnamefont {Kaufmann}}, \bibinfo
  {author} {\bibfnamefont {B.}~\bibnamefont {Pressl}}, \ and\ \bibinfo {author}
  {\bibfnamefont {G.}~\bibnamefont {Weihs}},\ }\href@noop {} {\bibfield
  {journal} {\bibinfo  {journal} {arXiv:1508.03253}\ } (\bibinfo {year}
  {2015})}\BibitemShut {NoStop}%
\bibitem [{\citenamefont {Metcalf}\ \emph {et~al.}(2013)\citenamefont
  {Metcalf}, \citenamefont {Thomas-Peter}, \citenamefont {Spring},
  \citenamefont {Kundys}, \citenamefont {Broome}, \citenamefont {Humphreys},
  \citenamefont {Jin}, \citenamefont {Barbieri}, \citenamefont {Kolthammer},
  \citenamefont {Gates}, \citenamefont {Smith}, \citenamefont {Langford},
  \citenamefont {Smith},\ and\ \citenamefont
  {Walmsley}}]{Metcalf:MultiPhotonInterferenceMultiport}%
  \BibitemOpen
  \bibfield  {author} {\bibinfo {author} {\bibfnamefont {B.~J.}\ \bibnamefont
  {Metcalf}}, \bibinfo {author} {\bibfnamefont {N.}~\bibnamefont
  {Thomas-Peter}}, \bibinfo {author} {\bibfnamefont {J.~B.}\ \bibnamefont
  {Spring}}, \bibinfo {author} {\bibfnamefont {D.}~\bibnamefont {Kundys}},
  \bibinfo {author} {\bibfnamefont {M.~A.}\ \bibnamefont {Broome}}, \bibinfo
  {author} {\bibfnamefont {P.~C.}\ \bibnamefont {Humphreys}}, \bibinfo {author}
  {\bibfnamefont {X.-M.}\ \bibnamefont {Jin}}, \bibinfo {author} {\bibfnamefont
  {M.}~\bibnamefont {Barbieri}}, \bibinfo {author} {\bibfnamefont {W.~S.}\
  \bibnamefont {Kolthammer}}, \bibinfo {author} {\bibfnamefont {J.~C.}\
  \bibnamefont {Gates}}, \bibinfo {author} {\bibfnamefont {B.~J.}\ \bibnamefont
  {Smith}}, \bibinfo {author} {\bibfnamefont {N.~K.}\ \bibnamefont {Langford}},
  \bibinfo {author} {\bibfnamefont {P.~G.}\ \bibnamefont {Smith}}, \ and\
  \bibinfo {author} {\bibfnamefont {I.~A.}\ \bibnamefont {Walmsley}},\ }\href
  {http://dx.doi.org/10.1038/ncomms2349} {\bibfield  {journal} {\bibinfo
  {journal} {Nat Commun}\ }\textbf {\bibinfo {volume} {4}},\ \bibinfo {pages}
  {1356} (\bibinfo {year} {2013})}\BibitemShut {NoStop}%
\bibitem [{\citenamefont {Carolan}\ \emph {et~al.}(2015)\citenamefont
  {Carolan}, \citenamefont {Harrold}, \citenamefont {Sparrow}, \citenamefont
  {Mart\'{i}n-L\'{o}pez}, \citenamefont {Russell}, \citenamefont {Silverstone},
  \citenamefont {Shadbolt}, \citenamefont {Matsuda}, \citenamefont {Oguma},
  \citenamefont {Itoh}, \citenamefont {Marshall}, \citenamefont {Thompson},
  \citenamefont {Matthews}, \citenamefont {Hashimoto}, \citenamefont
  {O'Brien},\ and\ \citenamefont {Laing}}]{Carolan:UniversalLinearOptics}%
  \BibitemOpen
  \bibfield  {author} {\bibinfo {author} {\bibfnamefont {J.}~\bibnamefont
  {Carolan}}, \bibinfo {author} {\bibfnamefont {C.}~\bibnamefont {Harrold}},
  \bibinfo {author} {\bibfnamefont {C.}~\bibnamefont {Sparrow}}, \bibinfo
  {author} {\bibfnamefont {E.}~\bibnamefont {Mart\'{i}n-L\'{o}pez}}, \bibinfo
  {author} {\bibfnamefont {N.~J.}\ \bibnamefont {Russell}}, \bibinfo {author}
  {\bibfnamefont {J.~W.}\ \bibnamefont {Silverstone}}, \bibinfo {author}
  {\bibfnamefont {P.~J.}\ \bibnamefont {Shadbolt}}, \bibinfo {author}
  {\bibfnamefont {N.}~\bibnamefont {Matsuda}}, \bibinfo {author} {\bibfnamefont
  {M.}~\bibnamefont {Oguma}}, \bibinfo {author} {\bibfnamefont
  {M.}~\bibnamefont {Itoh}}, \bibinfo {author} {\bibfnamefont {G.~D.}\
  \bibnamefont {Marshall}}, \bibinfo {author} {\bibfnamefont {M.~G.}\
  \bibnamefont {Thompson}}, \bibinfo {author} {\bibfnamefont {J.~C.~F.}\
  \bibnamefont {Matthews}}, \bibinfo {author} {\bibfnamefont {T.}~\bibnamefont
  {Hashimoto}}, \bibinfo {author} {\bibfnamefont {J.~L.}\ \bibnamefont
  {O'Brien}}, \ and\ \bibinfo {author} {\bibfnamefont {A.}~\bibnamefont
  {Laing}},\ }\href {http://www.sciencemag.org/content/349/6249/711.abstract}
  {\bibfield  {journal} {\bibinfo  {journal} {Science}\ }\textbf {\bibinfo
  {volume} {349}},\ \bibinfo {pages} {711} (\bibinfo {year}
  {2015})}\BibitemShut {NoStop}%
\bibitem [{\citenamefont {Vergyris}\ \emph {et~al.}(2016)\citenamefont
  {Vergyris}, \citenamefont {Meany}, \citenamefont {Lunghi}, \citenamefont
  {Downes}, \citenamefont {Steel}, \citenamefont {Withford}, \citenamefont
  {Alibart},\ and\ \citenamefont {Tanzilli}}]{Vergyris:OnChipHeraldedPairs}%
  \BibitemOpen
  \bibfield  {author} {\bibinfo {author} {\bibfnamefont {P.}~\bibnamefont
  {Vergyris}}, \bibinfo {author} {\bibfnamefont {T.}~\bibnamefont {Meany}},
  \bibinfo {author} {\bibfnamefont {T.}~\bibnamefont {Lunghi}}, \bibinfo
  {author} {\bibfnamefont {J.}~\bibnamefont {Downes}}, \bibinfo {author}
  {\bibfnamefont {M.~J.}\ \bibnamefont {Steel}}, \bibinfo {author}
  {\bibfnamefont {M.~J.}\ \bibnamefont {Withford}}, \bibinfo {author}
  {\bibfnamefont {O.}~\bibnamefont {Alibart}}, \ and\ \bibinfo {author}
  {\bibfnamefont {S.}~\bibnamefont {Tanzilli}},\ }\href@noop {} {\bibfield
  {journal} {\bibinfo  {journal} {arXiv:1605.03777}\ } (\bibinfo {year}
  {2016})}\BibitemShut {NoStop}%
\bibitem [{\citenamefont {Lee}\ \emph {et~al.}(2015)\citenamefont {Lee},
  \citenamefont {Moon}, \citenamefont {Choi},\ and\ \citenamefont
  {Park}}]{Lee:MCFSLM}%
  \BibitemOpen
  \bibfield  {author} {\bibinfo {author} {\bibfnamefont {H.~J.}\ \bibnamefont
  {Lee}}, \bibinfo {author} {\bibfnamefont {H.~S.}\ \bibnamefont {Moon}},
  \bibinfo {author} {\bibfnamefont {S.-K.}\ \bibnamefont {Choi}}, \ and\
  \bibinfo {author} {\bibfnamefont {H.~S.}\ \bibnamefont {Park}},\ }\href
  {\doibase 10.1364/OE.23.012555} {\bibfield  {journal} {\bibinfo  {journal}
  {Opt. Express}\ }\textbf {\bibinfo {volume} {23}},\ \bibinfo {pages} {12555}
  (\bibinfo {year} {2015})}\BibitemShut {NoStop}%
\bibitem [{\citenamefont {Kastner}(2016)}]{kastner16}%
  \BibitemOpen
  \bibfield  {author} {\bibinfo {author} {\bibfnamefont {R.~E.}\ \bibnamefont
  {Kastner}},\ }\href@noop {} {\bibfield  {journal} {\bibinfo  {journal}
  {arXiv:1603.05301}\ } (\bibinfo {year} {2016})}\BibitemShut {NoStop}%
\bibitem [{\citenamefont {Gr\"{a}fe}\ \emph {et~al.}(2012)\citenamefont
  {Gr\"{a}fe}, \citenamefont {Solntsev}, \citenamefont {Keil}, \citenamefont
  {Sukhorukov}, \citenamefont {Heinrich}, \citenamefont {T\"{u}nnermann},
  \citenamefont {Nolte}, \citenamefont {Szameit},\ and\ \citenamefont
  {Kivshar}}]{Graefe:2DSimulatorBiphotonWalk}%
  \BibitemOpen
  \bibfield  {author} {\bibinfo {author} {\bibfnamefont {M.}~\bibnamefont
  {Gr\"{a}fe}}, \bibinfo {author} {\bibfnamefont {A.~S.}\ \bibnamefont
  {Solntsev}}, \bibinfo {author} {\bibfnamefont {R.}~\bibnamefont {Keil}},
  \bibinfo {author} {\bibfnamefont {A.~A.}\ \bibnamefont {Sukhorukov}},
  \bibinfo {author} {\bibfnamefont {M.}~\bibnamefont {Heinrich}}, \bibinfo
  {author} {\bibfnamefont {A.}~\bibnamefont {T\"{u}nnermann}}, \bibinfo
  {author} {\bibfnamefont {S.}~\bibnamefont {Nolte}}, \bibinfo {author}
  {\bibfnamefont {A.}~\bibnamefont {Szameit}}, \ and\ \bibinfo {author}
  {\bibfnamefont {Y.~S.}\ \bibnamefont {Kivshar}},\ }\href@noop {} {\bibfield
  {journal} {\bibinfo  {journal} {Sci. Rep.}\ }\textbf {\bibinfo {volume}
  {2}},\ \bibinfo {pages} {562} (\bibinfo {year} {2012})}\BibitemShut {NoStop}%
\bibitem [{\citenamefont {Perez-Leija}\ \emph {et~al.}(2013)\citenamefont
  {Perez-Leija}, \citenamefont {Hernandez-Herrejon}, \citenamefont
  {Moya-Cessa}, \citenamefont {Szameit},\ and\ \citenamefont
  {Christodoulides}}]{Perez-Leija2013}%
  \BibitemOpen
  \bibfield  {author} {\bibinfo {author} {\bibfnamefont {A.}~\bibnamefont
  {Perez-Leija}}, \bibinfo {author} {\bibfnamefont {J.~C.}\ \bibnamefont
  {Hernandez-Herrejon}}, \bibinfo {author} {\bibfnamefont {H.}~\bibnamefont
  {Moya-Cessa}}, \bibinfo {author} {\bibfnamefont {A.}~\bibnamefont {Szameit}},
  \ and\ \bibinfo {author} {\bibfnamefont {D.~N.}\ \bibnamefont
  {Christodoulides}},\ }\href {\doibase 10.1103/PhysRevA.87.013842} {\bibfield
  {journal} {\bibinfo  {journal} {Phys. Rev. A}\ }\textbf {\bibinfo {volume}
  {87}},\ \bibinfo {pages} {013842} (\bibinfo {year} {2013})}\BibitemShut
  {NoStop}%
\bibitem [{\citenamefont {Meany}\ \emph {et~al.}(2015)\citenamefont {Meany},
  \citenamefont {Gr\"afe}, \citenamefont {Heilmann}, \citenamefont
  {Perez-Leija}, \citenamefont {Gross}, \citenamefont {Steel}, \citenamefont
  {Withford},\ and\ \citenamefont
  {Szameit}}]{Meany:LaserWrittenQuantumCircuitsReview}%
  \BibitemOpen
  \bibfield  {author} {\bibinfo {author} {\bibfnamefont {T.}~\bibnamefont
  {Meany}}, \bibinfo {author} {\bibfnamefont {M.}~\bibnamefont {Gr\"afe}},
  \bibinfo {author} {\bibfnamefont {R.}~\bibnamefont {Heilmann}}, \bibinfo
  {author} {\bibfnamefont {A.}~\bibnamefont {Perez-Leija}}, \bibinfo {author}
  {\bibfnamefont {S.}~\bibnamefont {Gross}}, \bibinfo {author} {\bibfnamefont
  {M.~J.}\ \bibnamefont {Steel}}, \bibinfo {author} {\bibfnamefont {M.~J.}\
  \bibnamefont {Withford}}, \ and\ \bibinfo {author} {\bibfnamefont
  {A.}~\bibnamefont {Szameit}},\ }\href {\doibase 10.1002/lpor.201500061}
  {\bibfield  {journal} {\bibinfo  {journal} {Laser Photonics Rev.}\ }\textbf
  {\bibinfo {volume} {9}},\ \bibinfo {pages} {363} (\bibinfo {year}
  {2015})}\BibitemShut {NoStop}%
\bibitem [{\citenamefont {Heilmann}\ and\ \citenamefont
  {Szameit}()}]{Heilmann:AdiabaticWriting}%
  \BibitemOpen
  \bibfield  {author} {\bibinfo {author} {\bibfnamefont {R.}~\bibnamefont
  {Heilmann}}\ and\ \bibinfo {author} {\bibfnamefont {A.}~\bibnamefont
  {Szameit}},\ }\href@noop {} {\bibinfo  {journal} {(to be published)}\
  }\BibitemShut {NoStop}%
\bibitem [{\citenamefont {Hunsperger}(1991)}]{Hunsperger:IntegratedOpticsBook}%
  \BibitemOpen
\bibfield  {journal} {  }\bibfield  {author} {\bibinfo {author} {\bibfnamefont
  {R.~G.}\ \bibnamefont {Hunsperger}},\ }\href {\doibase 10.1007/b98730} {\emph
  {\bibinfo {title} {Integrated optics}}}\ (\bibinfo  {publisher} {Springer
  (New York)},\ \bibinfo {year} {1991})\BibitemShut {NoStop}%
\bibitem [{\citenamefont {Grubbs}(1969)}]{grubbs69}%
  \BibitemOpen
  \bibfield  {author} {\bibinfo {author} {\bibfnamefont {F.~E.}\ \bibnamefont
  {Grubbs}},\ }\href {\doibase 10.1080/00401706.1969.10490657} {\bibfield
  {journal} {\bibinfo  {journal} {Technometrics}\ }\textbf {\bibinfo {volume}
  {11}},\ \bibinfo {pages} {1} (\bibinfo {year} {1969})}\BibitemShut {NoStop}%
\bibitem [{\citenamefont {Park}, \citenamefont {Moussa},\ and\ \citenamefont
  {Laflamme}(2012)}]{park12}%
  \BibitemOpen
  \bibfield  {author} {\bibinfo {author} {\bibfnamefont {D.~K.}\ \bibnamefont
  {Park}}, \bibinfo {author} {\bibfnamefont {O.}~\bibnamefont {Moussa}}, \ and\
  \bibinfo {author} {\bibfnamefont {R.}~\bibnamefont {Laflamme}},\ }\href
  {http://stacks.iop.org/1367-2630/14/i=11/a=113025} {\bibfield  {journal}
  {\bibinfo  {journal} {New J. Phys.}\ }\textbf {\bibinfo {volume} {14}},\
  \bibinfo {pages} {113025} (\bibinfo {year} {2012})}\BibitemShut {NoStop}%
\bibitem [{\citenamefont {Marcinkevicius}\ \emph {et~al.}(2001)\citenamefont
  {Marcinkevicius}, \citenamefont {Juodkazis}, \citenamefont {Watanabe},
  \citenamefont {Miwa}, \citenamefont {Matsuo}, \citenamefont {Misawa},\ and\
  \citenamefont {Nishii}}]{Marcinkevicius:LaserChannelEtching}%
  \BibitemOpen
  \bibfield  {author} {\bibinfo {author} {\bibfnamefont {A.}~\bibnamefont
  {Marcinkevicius}}, \bibinfo {author} {\bibfnamefont {S.}~\bibnamefont
  {Juodkazis}}, \bibinfo {author} {\bibfnamefont {M.}~\bibnamefont {Watanabe}},
  \bibinfo {author} {\bibfnamefont {M.}~\bibnamefont {Miwa}}, \bibinfo {author}
  {\bibfnamefont {S.}~\bibnamefont {Matsuo}}, \bibinfo {author} {\bibfnamefont
  {H.}~\bibnamefont {Misawa}}, \ and\ \bibinfo {author} {\bibfnamefont
  {J.}~\bibnamefont {Nishii}},\ }\href
  {http://ol.osa.org/abstract.cfm?URI=ol-26-5-277} {\bibfield  {journal}
  {\bibinfo  {journal} {Opt. Lett.}\ }\textbf {\bibinfo {volume} {26}},\
  \bibinfo {pages} {277} (\bibinfo {year} {2001})}\BibitemShut {NoStop}%
\bibitem [{\citenamefont {Crespi}\ \emph {et~al.}(2012)\citenamefont {Crespi},
  \citenamefont {Lobino}, \citenamefont {Matthews}, \citenamefont {Politi},
  \citenamefont {Neal}, \citenamefont {Ramponi}, \citenamefont {Osellame},\
  and\ \citenamefont {O'Brien}}]{Crespi:MeasuringProteinConc_EntPhot}%
  \BibitemOpen
  \bibfield  {author} {\bibinfo {author} {\bibfnamefont {A.}~\bibnamefont
  {Crespi}}, \bibinfo {author} {\bibfnamefont {M.}~\bibnamefont {Lobino}},
  \bibinfo {author} {\bibfnamefont {J.~C.~F.}\ \bibnamefont {Matthews}},
  \bibinfo {author} {\bibfnamefont {A.}~\bibnamefont {Politi}}, \bibinfo
  {author} {\bibfnamefont {C.~R.}\ \bibnamefont {Neal}}, \bibinfo {author}
  {\bibfnamefont {R.}~\bibnamefont {Ramponi}}, \bibinfo {author} {\bibfnamefont
  {R.}~\bibnamefont {Osellame}}, \ and\ \bibinfo {author} {\bibfnamefont
  {J.~L.}\ \bibnamefont {O'Brien}},\ }\href {\doibase
  http://dx.doi.org/10.1063/1.4724105} {\bibfield  {journal} {\bibinfo
  {journal} {Appl. Phys. Lett.}\ }\textbf {\bibinfo {volume} {100}},\ \bibinfo
  {eid} {233704} (\bibinfo {year} {2012})}\BibitemShut {NoStop}%
\bibitem [{\citenamefont {Pernice}\ \emph {et~al.}(2012)\citenamefont
  {Pernice}, \citenamefont {Schuck}, \citenamefont {Minaeva}, \citenamefont
  {Li}, \citenamefont {Goltsman}, \citenamefont {Sergienko},\ and\
  \citenamefont {Tang}}]{Pernice:SCNanowiresOnChip}%
  \BibitemOpen
  \bibfield  {author} {\bibinfo {author} {\bibfnamefont {W.}~\bibnamefont
  {Pernice}}, \bibinfo {author} {\bibfnamefont {C.}~\bibnamefont {Schuck}},
  \bibinfo {author} {\bibfnamefont {O.}~\bibnamefont {Minaeva}}, \bibinfo
  {author} {\bibfnamefont {M.}~\bibnamefont {Li}}, \bibinfo {author}
  {\bibfnamefont {G.}~\bibnamefont {Goltsman}}, \bibinfo {author}
  {\bibfnamefont {A.}~\bibnamefont {Sergienko}}, \ and\ \bibinfo {author}
  {\bibfnamefont {H.}~\bibnamefont {Tang}},\ }\href
  {http://dx.doi.org/10.1038/ncomms2307} {\bibfield  {journal} {\bibinfo
  {journal} {Nat Commun}\ }\textbf {\bibinfo {volume} {3}},\ \bibinfo {pages}
  {1325} (\bibinfo {year} {2012})}\BibitemShut {NoStop}%
\bibitem [{\citenamefont {Kauten}\ \emph {et~al.}(2014)\citenamefont {Kauten},
  \citenamefont {Pressl}, \citenamefont {Kaufmann},\ and\ \citenamefont
  {Weihs}}]{kauten14}%
  \BibitemOpen
  \bibfield  {author} {\bibinfo {author} {\bibfnamefont {T.}~\bibnamefont
  {Kauten}}, \bibinfo {author} {\bibfnamefont {B.}~\bibnamefont {Pressl}},
  \bibinfo {author} {\bibfnamefont {T.}~\bibnamefont {Kaufmann}}, \ and\
  \bibinfo {author} {\bibfnamefont {G.}~\bibnamefont {Weihs}},\ }\href
  {http://scitation.aip.org/content/aip/journal/rsi/85/6/10.1063/1.4879820}
  {\bibfield  {journal} {\bibinfo  {journal} {Rev. Sci. Instrum.}\ }\textbf
  {\bibinfo {volume} {85}},\ \bibinfo {eid} {063102} (\bibinfo {year}
  {2014})}\BibitemShut {NoStop}%
\bibitem [{\citenamefont {Procopio}\ \emph {et~al.}(2016)\citenamefont
  {Procopio}, \citenamefont {Rozema}, \citenamefont {Wong}, \citenamefont
  {Hamel}, \citenamefont {O'Brien}, \citenamefont {Zhang}, \citenamefont
  {Daki\'{c}},\ and\ \citenamefont
  {Walther}}]{Procopio:SagnacHypercomplexTest}%
  \BibitemOpen
  \bibfield  {author} {\bibinfo {author} {\bibfnamefont {L.~M.}\ \bibnamefont
  {Procopio}}, \bibinfo {author} {\bibfnamefont {L.~A.}\ \bibnamefont
  {Rozema}}, \bibinfo {author} {\bibfnamefont {Z.~J.}\ \bibnamefont {Wong}},
  \bibinfo {author} {\bibfnamefont {D.~R.}\ \bibnamefont {Hamel}}, \bibinfo
  {author} {\bibfnamefont {K.}~\bibnamefont {O'Brien}}, \bibinfo {author}
  {\bibfnamefont {X.}~\bibnamefont {Zhang}}, \bibinfo {author} {\bibfnamefont
  {B.}~\bibnamefont {Daki\'{c}}}, \ and\ \bibinfo {author} {\bibfnamefont
  {P.}~\bibnamefont {Walther}},\ }\href@noop {} {\bibfield  {journal} {\bibinfo
   {journal} {arXiv:1602.01624}\ } (\bibinfo {year} {2016})}\BibitemShut
  {NoStop}%
\end{thebibliography}

%

\end{document}